\documentstyle[prl,aps,epsf,amsfonts,amssymb,multicol,epsfig]{revtex}
\begin{document}
\title{Effects of strain, electric, and magnetic fields
  on lateral electron spin transport in semiconductor epilayers}
\author {M. Hru\v{s}ka $^1$, \v{S}. Kos $^{1,2}$, S. A. Crooker $^3$,
 A. Saxena $^1$ and D. L. Smith $^1$}
\address{$^1$Theoretical Division, Los Alamos National Laboratory,
Los Alamos, New Mexico 87545, USA \\
 $^2$Cavendish Laboratory, Cambridge University, Madingley Road, Cambridge,
 CB3 0HE, United Kingdom \\
  $^3$National High Magnetic Field Laboratory, Los Alamos National Laboratory,
 Los Alamos, New Mexico 87545, USA }
\maketitle

\begin{abstract}
We construct a spin-drift-diffusion model to describe
spin-polarized electron transport in zincblende semiconductors in
the presence of magnetic fields, electric fields, and off-diagonal
strain. We present predictions of the model for geometries that
correspond to optical spin injection from the absorption of
circularly polarized light, and for geometries that correspond to
electrical spin injection from ferromagnetic contacts.  Starting
with the Keldysh Green's function description for a system driven
out of equilibrium, we construct a semiclassical kinetic theory of
electron spin transport in strained semiconductors in the presence
of electric and magnetic fields.  From this kinetic theory we
derive spin-drift-diffusion equations for the components of the
spin density matrix for the specific case of spatially uniform
fields and uniform electron density.  We solve the
spin-drift-diffusion equations numerically and compare the
resulting images with scanning Kerr microscopy data of
spin-polarized conduction electrons flowing laterally in bulk
epilayers of n-type GaAs. The spin-drift-diffusion model
accurately describes the experimental observations.  We contrast
the properties of electron spin precession resulting from magnetic
and strain fields. Spin-strain coupling depends linearly on
electron wave vector and spin-magnetic field coupling is
independent of electron wave vector. As a result, spatial
coherence of precessing spin flows is better maintained with
strain than with magnetic fields, and the spatial period of spin
precession is independent of the applied electrical bias in
strained structures whereas it is strongly bias dependent for the case of 
applied magnetic fields.

PACS numbers: 72.25.Dc, 71.70.Ej, 85.75.-d
\end{abstract}

\begin{multicols}{2}

\section{Introduction}

A new generation of electronic devices, with the potential to
outperform conventional electronic circuits in speed, integration
density and power consumption, has been proposed based on the
ability to manipulate electron spin in semiconductors
\cite{reviews,DattaDas,newerspintransistor}. To design
semiconductor structures whose function is based on electron spin,
it is necessary to understand spin dynamics and spin-polarized
transport, and in particular, how they are affected by electric,
magnetic and strain fields. Spin dynamics and spin transport in
semiconductors have been studied experimentally using time- and/or
spatially-resolved spin-sensitive optical spectroscopies based on
the magneto-optical Faraday and Kerr effects
\cite{DDA,CrookerIEEE,Crooker,Kikkawa99,Frey,Kimel,Sandhu,Stephens,Sanada,Nature,KatoPRL,ScottDarryl,science,Beck}.
Theoretical approaches to describe spin dynamics and
spin-polarized transport include: the two-component
drift-diffusion model\cite{Flatte,Yu,Zutic,PershinPrivman},
Boltzmann equation approaches
\cite{Weng1,Weng2,Weng3,Fabian,Ivar,Bronold,Kalevich,ValetFert,Grimaldi,Bleibaum},
Monte-Carlo techniques \cite{Kiselev,Pramanik,Pershin} and
microscopic approaches \cite{Mishchenko,Schmeltzer,Nikolic}.  Many
of these works have focused on the description of spin-relaxation
times.  In this paper we are particularly interested in describing
the effects of strain on spin-polarized electron transport.

Recently, the possibility of using strain to control electron spin
precession in zincblende structure semiconductors has been
demonstrated \cite{Nature,KatoPRL,ScottDarryl,science,Beck}. The
spatial part of the conduction electron wave function is modified
in the presence of strain which affects the electron spin degrees
of freedom due to spin-orbit coupling.  Scanning polar
Kerr-rotation microscopy (in which a linearly polarized laser
propagating along the sample normal is raster scanned in the x-y
sample plane and the polarization rotation angle of the reflected
beam is detected) has been shown to be a valuable experimental
technique to image electron spin flow in zincblende structure
semiconductors. Scanning Kerr microscopy has recently been used to
acquire 2-D images of electron spin flows in n-type GaAs epilayers
subject to uniform electric, magnetic, and (off-diagonal) strain
fields in which spin-polarized electrons have been optically
injected by circularly polarized light \cite{ScottDarryl} or
electrically injected into lateral spin-transport devices using
ferromagnetic iron (Fe) contacts \cite{science}.

In this paper, we microscopically derive an equation of motion for
the electron Green's function giving a full quantum-mechanical
description of electron spin dynamics and transport in the
presence of electric, magnetic and strain fields.  From this
equation of motion we construct a semiclassical kinetic theory of
electron spin dynamics and transport in the presence of these
fields. From the semiclassical kinetic theory, we derive a set of
spin-drift-diffusion equations for the components of the spin
density matrix for the case of spatially uniform fields and
electron density.  This approach extends beyond the two-component
drift-diffusion model, because the spin distribution function has
the form of a spin-density matrix to account for spin coherence
effects. We are particularly interested in describing spin
polarized electron transport in the presence of momentum-dependent
coupling between spin and off-diagonal strain.  Off-diagonal
strain arises, e.g., from uniaxial stress along a
$\langle110\rangle$ crystal axis. Because the spin-strain coupling
depends on electron wave vector, the diffusion coefficient and
mobility appear in the strain coupling terms of the
spin-drift-diffusion equation. We solve the spin-drift-diffusion
equations numerically and find good agreement with scanning Kerr
microscopy images of spin-polarized conduction electrons flowing
laterally in bulk epilayers of n-type GaAs.  We contrast the
effects of magnetic and strain fields on electron spin transport.
Both magnetic fields and strain fields (with off-diagonal strain
components) lead to electron spin precession in zincblende
semiconductor structures. However, spin-strain coupling is linear
in electron wave vector whereas spin-magnetic field coupling is
independent of electron wave vector, leading to qualitatively
different behavior. We present calculations that contrast spin
transport in strain and magnetic fields and then present a series
of theoretical predictions for (and calculated images of) spin
transport in geometries that correspond to optical spin injection
and in geometries that correspond to electrical spin injection
into lateral devices.

The outline of the paper is as follows: in Sec. II, we solve the
spin-drift-diffusion equations for the spin density matrix
(derived in the Appendix) and we compare results with experimental
data; in Sec. III, we discuss differences between strain-induced
and magnetic-field-induced spin precession; a series of
theoretical predictions for spin transport are presented in Sec.
IV; and in Sec. V, we summarize our conclusions.  Details of the
derivation of the transport equations are presented in the
Appendix.

 \section{Spin-Drift-diffusion Equations}

Figure 1 shows schematics of the two experimental geometries
considered.  A [001]-oriented n:GaAs epilayer sample, whose
surface normal is along the z-axis, is subject to in-plane
electric and magnetic fields, as well as strain fields. As
illustrated in Fig. 1a, the sample may be optically excited by a
circularly polarized laser beam propagating along the surface
normal so that electrons, spin polarized along the z-axis, are
optically injected. Alternatively, spin polarized electrons may be
electrically injected into the semiconductor from ferromagnetic
(FM) contacts as illustrated in Fig. 1b. For electrical injection,
the injected electron spin polarization follows the magnetization
{\bf M} of the ferromagnetic contact, and is therefore typically
in the x-y plane of the sample surface (for certain special
ferromagnetic contacts both {\bf M} and the injected electron spin
polarization may be directed along the surface normal). The spin
polarized electrons subsequently drift and diffuse in the x-y
sample plane. The epilayer, in which the electrons are confined,
is thin compared to a spin diffusion length so that the spin
density is essentially uniform in the z-direction.  The resulting
steady state spin polarization is imaged via scanning Kerr
microscopy, which is sensitive to the z-component of electron spin
polarization, $n_z$. In the experiments
\cite{ScottDarryl,science}, ohmic side contacts allow for a
lateral electrical bias ($E_x$) in the x-y plane, Helmholtz coils
provide an in-plane magnetic field ($B_y$), and uniform
off-diagonal strain ($\epsilon_{xy}$) is provided by a controlled
uniaxial stress applied to the sample along a [110] crystal axis.
The experimental data in the lower panels of Fig. 1 show false
color 2-D images of the measured z-component of electron spin
polarization $n_z$, showing the flow of spin polarized electrons.
In the lower left panel, spin-polarized electrons are optically
injected using a laser focused to a 4 $\mu$m spot (spot size
illustrated by the circular red dot) and these spins subsequently
diffuse and drift to the right in an applied electric field ($E_x
= 10$ V/cm). In the lower right panel spin-polarized electrons are
electrically injected from an iron (Fe) tunnel-barrier contact
that is magnetized along $-\hat{x}$. The injected spin
polarization is in the x-direction, and a small magnetic field in
the y-direction ($B_y=3.6$ G) rotates the spins so that there is a
z-component of spin polarization $n_z$ that can be detected by
Kerr microscopy.

\begin{figure}
\begin{center}
\epsfig{file=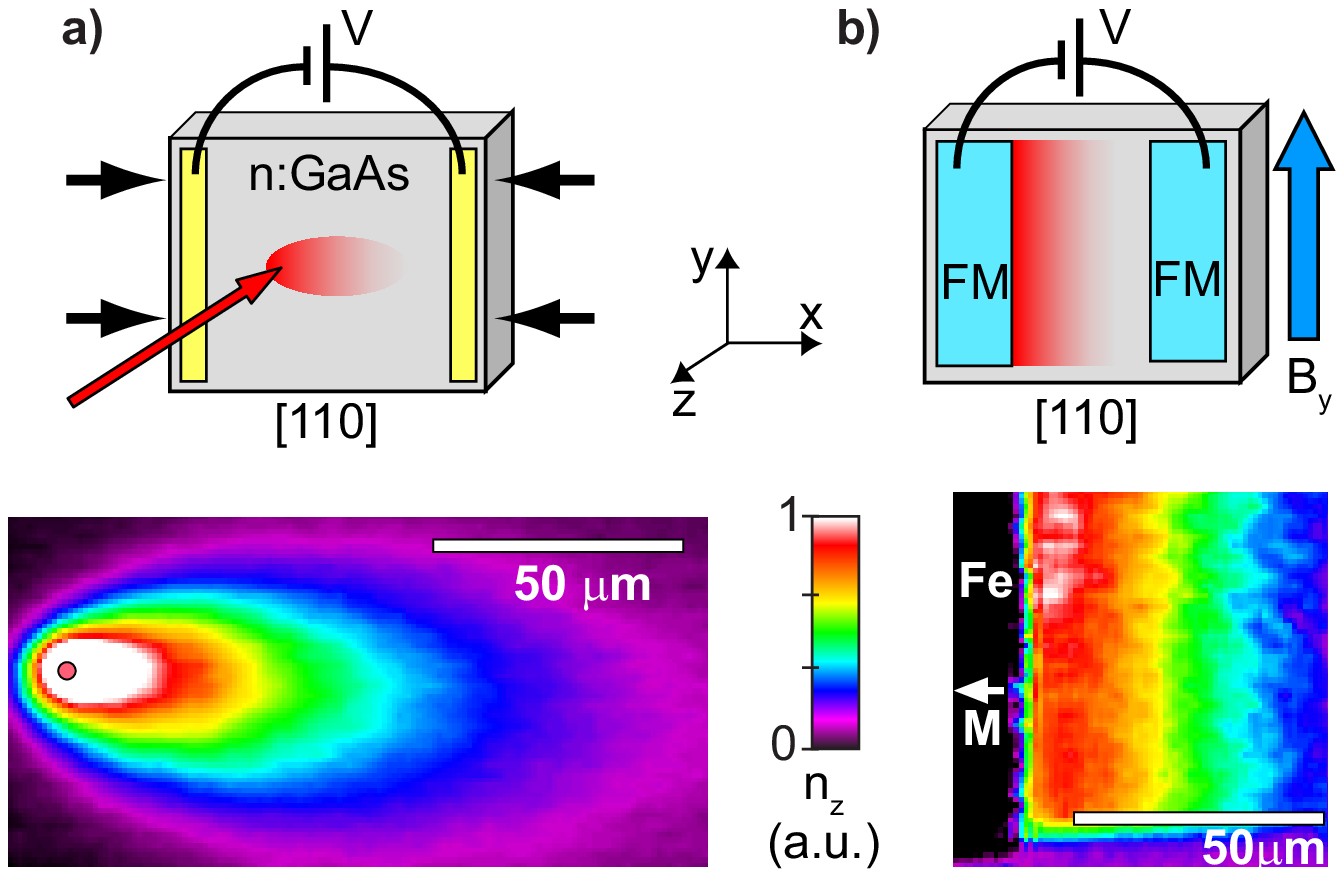,width=0.5\textwidth,clip=}
\end{center}
\caption{Diagrams and corresponding experimental data of the two
measurement geometries considered in this work. (a) Spin polarized
electrons, oriented along $\hat{z}$, are optically injected into
an n:GaAs epilayer with a focused, circularly polarized laser
incident along the z-axis (indicated by the red arrow). Lateral
contacts provide an in-plane electric field (${\bf E} \parallel
\pm \hat{x}$), Helmholtz coils provide an in-plane magnetic field
(${\bf B} \parallel \pm \hat{y}$), and off-diagonal strain
$\epsilon_{xy}$ results from controlled uniaxial stress applied
along the [110] GaAs crystal axis. Scanning polar Kerr-rotation
microscopy is used to spatially image $n_z$, the z-component of
the resulting steady-state electron spin polarization. The drift,
diffusion and precession of the spin flow are determined (in part)
by the applied electric, magnetic and strain fields (see data
below; adapted from Ref. [15]). 
(b) Electrical
injection of spin-polarized electrons into an n:GaAs epilayer
using ferromagnetic (FM) injector contacts. The contact
magnetization ({\bf M}) may be either in the x-y plane, or
out-of-plane ({\bf M}$\|\hat{z}$).  As shown schematically, the
injected spins then flow laterally within the n:GaAs epilayer, and
$n_z$ can be spatially imaged with scanning Kerr microscopy.
Experimental data, using an iron injector with {\bf M}$\|-\hat{x}$
and $B_y=3.6$ G is shown below (adapted from Ref. [16]). 
\label{Fig1}}
\end{figure}

To theoretically describe the effects of spin-strain coupling, we
consider a Hamiltonian of the form
\begin{eqnarray}
&H= \frac{1}{2 m} [ {\bf p} -\frac{e}{c} {\bf A} ({\bf R}) ]^2 +e
\phi ({\bf R}) + \frac{\hbar}{2}  \mbox {\boldmath $\Omega$}^{eff}
\cdot \mbox {\boldmath $\sigma$},&
\\
& \mbox {$\hbar$ \boldmath $\Omega$} ^{eff}=  g \mu _B {\bf B} +
C_3 \mbox {\boldmath $ \varphi$}  \ .& \nonumber
\end{eqnarray}
(Details of our microscopic treatment, that also include
Dresselhaus \cite{Dresselhaus} and Rashba \cite{Rashba} spin
coupling terms, are described in the Appendix.) Here $\mbox
{\boldmath $\sigma$}$ is the vector having the three Pauli spin
matrices as its components, $e=-|e|$ is the charge of an electron,
$m$ is the conduction electron effective mass, $\phi$ is the
scalar potential, $g$ is the conduction electron gyromagnetic
ratio, $\mu _B$ is the Bohr magneton, $\bf B$ is the applied
magnetic field and ${\bf A}$ is the corresponding vector
potential. The term
\begin{equation}
\varphi _x = (\epsilon _{xy} k_y - \epsilon _{xz} k_z),
\end{equation}
where $\varphi_y$ and $\varphi_z$ are obtained by cyclic
permutations, describes the lowest-order $\bf k$-linear coupling
of electron spin to the off-diagonal components of the crystal
strain tensor\cite{Meier}. Here, $\epsilon _{ij}$ is the strain
tensor,
\[
C_3 = \frac{4}{3} \hbar \frac{C_2 \eta }{ \sqrt{ 2 E_g m (1- \eta
/3)}} \ , \ \ \ \eta=\frac{\Delta}{E_g + \Delta} \ ,
\]
$C_2$ is the interband deformation potential constant for acoustic
phonons, $\Delta$ is the spin-orbit splitting of the valence band
and $E_g$ is the band gap energy. For electrons moving laterally
in the x-y sample plane, the spin-strain coupling term takes the
form $\epsilon_{xy} (\sigma_x k_y - \sigma_y k_x)$, which has the
same form as the Rashba spin coupling in asymmetric
heterostructures. That is, the presence of off-diagonal strain
results in an effective magnetic field {\bf B}$_\epsilon$ which is
in-plane, orthogonal to the electron momentum {\bf k}, and grows
linearly with the magnitude of {\bf k}. In contrast, spin-magnetic
field coupling is independent of {\bf k}.  Because of this
difference, strain and magnetic fields affect the flow of
spin-polarized electrons in qualitatively different ways.

We derive spin-drift-diffusion equations for the spin density
matrix
\begin{equation}
 \rho _s = n ({\bf R}) + \sum _{i=x,y,z} n_i ({\bf R}) \sigma _i ,
\end{equation}
accounting for the quantum nature of spin.  Here $n$ is the total
electron density and $n_i$ denote the three spin density
components.  Information about spin polarization along any given
direction $\hat s$ is found by Tr$\rho _s \sum_{i=x,y,z} s_i
\sigma _i $.  We consider a case, corresponding to the
experimental geometry, in which the net electron density is
position independent and the electric, magnetic and strain fields
are uniform. As shown in the Appendix, for electric and magnetic
fields in the x-y plane and stress along a [110] crystal axis, the
system of spin-drift-diffusion equations for the spin density
matrix is given by
 \begin{eqnarray}
\left(D \nabla _{\bf R}^2 + \mu {\bf E}  \cdot  \nabla _{\bf R}
 -C_s^2 D - \frac{1}{\tau _s}\right)  n_x &&
\nonumber\\
-\left[-C_{B_y} -C_s \left(2 D \frac{\partial}{\partial x} + \mu
E_x \right)\right] n_z & =& -G_x({\bf R}) ,
\end{eqnarray}

 \begin{eqnarray}
\left(D \nabla _{\bf R}^2 + \mu {\bf E}  \cdot  \nabla _{\bf R}
  -C_s^2 D - \frac{1}{\tau _s}\right) n_y \ \ \ &&
\nonumber\\
-\left[C_{B_x} - C_s \left( 2D \frac{\partial}{\partial y} + \mu
E_y \right)\right] n_z
  &=& -G_y({\bf R}) ,
  \end{eqnarray}

  \begin{eqnarray}
\left(D \nabla _{\bf R}^2 + \mu {\bf E}  \cdot  \nabla _{\bf R}
  -2 C_s^2 D - \frac{1}{\tau _s}\right) &n_z &
\nonumber\\
+\left[-C_{B_y} - C_s \left(2D \frac{\partial}{\partial x}+\mu E_x
\right)\right]n_x &
\nonumber\\
+\left[C_{B_x} - C_s \left(2D \frac{\partial}{\partial y} + \mu
E_y \right)\right]n_y
  &=& -G_z({\bf R}) ,
 \end{eqnarray}
where $\tau _s$ is the spin relaxation time, $D$ is the electron
diffusion coefficient, $\mu$ is the electron mobility, $C_s=C_3 m
\epsilon_{xy} / \hbar ^2 $, $\hbar C_{B_i}= g \mu _B B_{i}$
($i=x,y$) and the vector $\bf G$ describes generation of the spin
components. Because the spin-strain coupling depends on electron
wave vector, both $D$ and $\mu$ appear in the strain coupling
terms. $D$ and $\mu$ do not appear in the magnetic field coupling
terms.  A second order term in strain, $C_s^2 D$, appears in the
drift diffusion equations. The analogous second order term in
magnetic field is proportional to the momentum relaxation time,
which is short on the time scales of interest here, and thus this
term is very small and can be neglected (see the Appendix for
details.)

Performing a 2-D Fourier transform to momentum space, the system
of equations for the spin polarization components, in our
experimental geometry, is given by
\begin{eqnarray}
 O_1({\bf q}) n_x({\bf q}) +  O_2({\bf q}) n_z({\bf q}) &=&  - G_x({\bf q}) \ ,
 \label{driftdiffx} \\
 O_1({\bf q}) n_y({\bf q}) +  O_3({\bf q}) n_z({\bf q}) &=&  - G_y({\bf q}) \ ,
\label{driftdiffy} \\
 O_4({\bf q}) n_z({\bf q}) -  O_2({\bf q}) n_x({\bf q})
 - O_3({\bf q}) n_y({\bf q})&=& - G_z({\bf q}),
\label{driftdiffz}
\end{eqnarray}
where
\begin{eqnarray}
 O_1({\bf q}) &=& \left[-D q^2 + i \mu {\bf E} \cdot {\bf q } - C_s^2 D
- \frac{1}{\tau _s} \right] \,
 \\
 O_2({\bf q}) &=&  [C_{B_y} +C_s (i 2 D q_x + \mu E_x )]  \ ,
 \\
 O_3({\bf q}) &=& -[C_{B_x} -C_s (i 2 D q_y + \mu E_y )] \ ,
 \\
 O_4({\bf q}) &=& \left[-D q^2 + i \mu {\bf E} \cdot {\bf q } - 2 C_s^2 D
- \frac{1}{\tau _s} \right].
\end{eqnarray}
The resulting spin density, $n_i({\bf R})$, is then given by the
2-D Fourier transform of $n_i({\bf q})$,
\begin{eqnarray}
n_i({\bf R}) = \frac{1}{2 \pi} \int e^{i{\bf q} \cdot {\bf R}}
n_i({\bf q}) d^2 {\bf q}.
\end{eqnarray}

The solutions for the spin polarization components from this
system of equations are
\begin{eqnarray}
n_x({\bf q})= \frac{-(O_4 + \frac{{O_3}^2}{O_1})G_x + (\frac{O_2
O_3}{O_1})G_y + O_2 G_z}{d} ,
\end{eqnarray}
\begin{eqnarray}
n_y({\bf q})= \frac{(\frac{O_2 O_3}{O_1})G_x -(O_4 +
\frac{{O_2}^2}{O_1})G_y + O_3 G_z}{d} ,
\end{eqnarray}
\begin{eqnarray}
n_z({\bf q})= -\frac{O_2 G_x + O_3 G_y + O_1 G_z}{d} ,
\end{eqnarray}
where the denominator, $d$, is
\begin{eqnarray}
d=O_1 O_4 + (O_2)^2 + (O_3)^2.
\end{eqnarray}

In Fig. 2 we compare our calculations with experimental results
for the case of optical spin injection. These experimental data
were obtained via scanning Kerr microscopy \cite{ScottDarryl}. In
these measurements, a linearly polarized narrowband Ti:sapphire
laser, tuned just below the GaAs band-edge and focused to a 4
$\mu$m spot on the sample surface, was raster-scanned in the x-y
sample plane to construct a 2-D image of $n_z$, the z-component of
the electron spin density. The sample was a Si-doped n-type GaAs
epilayer (electron density = $1 \times 10^{16}$ cm$^{-3}$) grown
on a [001] oriented semi-insulating GaAs substrate. Spin polarized
electrons were optically injected into the sample by a separate,
circularly-polarized 1.58 eV diode laser that was also focused to
a 4 $\mu$m spot on the sample. Measurements were performed at a
temperature of 4 K. For additional experimental details, see
Ref.\cite{ScottDarryl}. Figures (a), (c), and (e) compare
calculated and measured results for spin flowing to the right in
the presence of uniform off-diagonal strain ($\epsilon_{xy} = 4
\times 10^{-4}$; $E_x = 12$ V/cm), while Figs. (b), (d), and (f)
compare calculated and measured results for spins flowing to the
right in the presence of an applied magnetic field ($B_y = 16$ G;
$E_x = 7$ V/cm). There is very good agreement between theory and
experiment. The material parameters used in the calculation were:
mobility $\mu=3000$ cm$^2$/Vs, spin relaxation time $\tau_s=125$
ns (we use these values for $\mu$ and $\tau_s$ throughout the
paper), and diffusion constant $D=10$ cm$^2$/s. Throughout the
paper we use a value of $C_3 = 4.0$ eV\AA~ for the spin-strain
coupling coefficient based on the experiments of Ref. \cite
{Cardona}.  Both strain and magnetic field lead to precession of
the electron spins. However, the spatial damping of the precession
is more pronounced when magnetic rather than strain fields are
applied (at the same precession length).  In the remainder of the 
paper, we discuss various predictions of the spin-drift-diffusion model.
\begin{figure}
\begin{center}
\epsfig{file=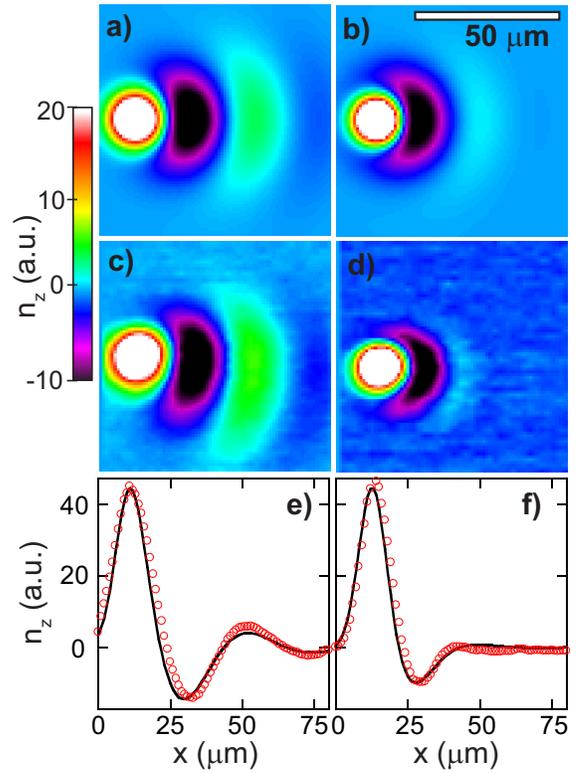,width=0.45\textwidth,clip=}
\end{center}
\caption{(a) and (b) show images of the calculated z-component of
electron spin density ($n_z$) for optically injected spins that
are flowing to the right and that are subject to (a) applied [110]
stress (strain $\epsilon_{xy}= 4 \times 10^{-4}$; $E_x = 12$
V/cm), and (b) applied magnetic field ($B_y = 16$ G; $E_x = 7$
V/cm). Note that the color scale is modified such that black
indicates negative values of $n_z$ (i.e., spins that have
precessed {\it into} the sample plane). For all calculations,
$D=10$ cm$^2$/s, $\mu=3000$ cm$^2$/Vs, and $\tau_s=125$ ns.  (c)
and (d) show images of corresponding experimental data under the
same conditions. (e) and (f) directly compare line cuts through
the center of the corresponding calculated (black line) and
measured (red dots) images.
 \label{Fig2}}
\end{figure}

\section{Contrasting the effects of magnetic and strain fields}

Although strain and magnetic fields both lead to electron spin
precession, spin precession due to strain differs qualitatively
from that due to a magnetic field because spin-strain coupling
depends linearly on the electron wave vector ${\bf k}$ whereas
spin-magnetic field coupling is independent of electron wave
vector.  For spin precession in an applied magnetic field there is
a pronounced spatial dephasing of spin precession due to the
randomizing nature of diffusion.  The net spin polarization at a
remote location from the point of spin generation results from the
combined sum of many random walk paths.  Each path takes a
different amount of time, giving a different precession angle. As
a result the spatial coherence of the spin flow is rapidly lost.
By contrast, the spin precession frequency in a strain field
scales linearly with the electron wave vector.  As a result the
precession angle is correlated with electron velocity and
therefore with the electron's position. There is a partial
cancelation of accumulated precession angle during the parts of
the diffusive path that are traversed in the opposite direction,
due to the linear dependence of effective magnetic field on
electron momentum. Indeed, for motion strictly in one dimension,
the spin ensemble would not dephase at all (however, it would
still decohere with a timescale $\tau_s$). Scattering from
$+\textbf{k}$ to $-\textbf{k}$ simply reverses the direction of
precession, leading to an exact correspondence between spatial
position and spin orientation.

A second difference between spin precession due to strain and that
due to magnetic field is that for spin-magnetic field coupling the
spatial precession length increases with increasing electric field
(drift velocity), whereas for spin-strain coupling this length
scale is independent of electric field. For spin-magnetic field
coupling, the spin precession frequency is independent of electric
field, but because the electrons drift faster in a larger electric
field, the spatial precession length increases.  For spin-strain
coupling, the precession frequency scales directly with the
average electron velocity, so that the spatial spin precession
length is nearly independent of electric field.

To illustrate these differences, we consider the simple
geometrical case in which spins are generated by a stripe optical
excitation along the y-axis, the electric field ${\bf E} \parallel
\hat{x}$, the magnetic field ${\bf B} \parallel \hat{y}$ and
off-diagonal strain $\epsilon_{xy}$ is generated by stress applied
along $\hat{x}$ (a [110] crystal axis). For this geometry,
spin-drift-diffusion occurs along the x-axis so that the spin
densities do not depend on the y-coordinate. Because of the
orientation of strain and magnetic field, $n_y$ vanishes. For this
geometrical case, $O_3$ vanishes and only the x-component, $q_x$,
is of interest. The Green's function for the z-component of spin
density precesses and decays exponentially as $g_z(x)=A \exp(\eta
\frac{x}{\sqrt{D \tau_s}})+CC$. The four complex decay parameters,
$\eta$, can be found from the four zeros of the denominator $d$ in
Eq. (18), $\eta = i q_x \sqrt{D \tau_s}$, where $q_x$ is a zero of
$d$. Setting $\alpha = {\mu E_x \tau_s}/{\sqrt{D\tau_s}}$,
$\beta_B = C_{B_y} \tau_s $, $\beta_s = C_s \sqrt{D \tau_s}$,
$Q=(\frac{\alpha}{2})^2 +1$, and $\zeta = \eta +
\frac{\alpha}{2}$, the zeros of $d$ are given by
\begin{eqnarray}
[{\zeta}^2 - (Q+{\beta_s}^2)] [{\zeta}^2 - (Q+2{\beta_s}^2)] =
-[-\beta_B + 2 \beta_s \zeta]^2. \nonumber\\
\end{eqnarray}

For {\it strictly} 1-D motion, in which the electrons are
constrained to move only along the x-axis, the
spin-drift-diffusion equations are modified in that the factor of
2 multiplying ${C_s}^2D$ in Eq. (6) is replaced by unity [see Eq.
(48)]. The corresponding factor of 2 that appears in $O_4(\bf q)$,
Eq. (13), and that multiplies ${\beta_s}^2$ in the second factor
on the left hand side of Eq. (19) are also replaced by unity for
this strictly 1-D case.  The magnetic field terms of the
spin-drift-diffusion equations are the same for the 1-D and 2-D
cases.

We consider the solutions of Eq. (19) for three cases: (1) applied
magnetic field without strain; (2) strain without magnetic field;
and (3) strain without magnetic field for strictly 1-D motion. For
case (1) the solutions are,
\begin{eqnarray}
\zeta = \pm \sqrt {\frac {\sqrt{Q^2 + {\beta_B}^2}+ Q}{2}} \pm i
\sqrt{\frac {\sqrt{Q^2 + {\beta_B}^2}- Q}{2}}.
\end{eqnarray}
For case (2) the solutions are,
\begin{eqnarray}
\zeta = \pm \sqrt {\frac {\sqrt{Q+{\beta_s}^2}
\sqrt{Q+2{\beta_s}^2} + (Q - \frac {{\beta_s}^2}{2})}{2}} &
\nonumber\\
\pm i \sqrt {\frac {\sqrt{Q+{\beta_s}^2} \sqrt{Q+2{\beta_s}^2} -
(Q - \frac {{\beta_s}^2}{2})}{2}}.
\end{eqnarray}
For case (3) the solutions are,
\begin{eqnarray}
\zeta = \pm \sqrt{Q} \pm i \beta_s.
\end{eqnarray}
From Eq. (20) we see that the real part of $\zeta$ increases with
increasing magnetic field so that the decay length for spin
density decreases with increasing magnetic field. From Eq. (21),
we see that the real part of $\zeta$ also increases with
increasing strain, but that there is a partial cancellation, due
to the appearance of the term $-\frac {{\beta_s}^2}{2}$.  The
decay length for spin polarization also decreases with increasing
strain, but more slowly than for magnetic field. To compare the
strain and magnetic field cases, we consider strains and magnetic
fields that give the same spatial precession length.  For the case
of strain in strictly 1-D motion, the factor of 2 and of $1/2$
that appear in the numerator of Eq. (21) are both replaced by
unity and the partial cancelations that occur for the 2-D case
become complete in the 1-D case, giving the result of Eq. (22). In
this case strain does not reduce the decay length of spin
polarization.

In Fig. 3, we calculate $n_z$ as a function of position for a
narrow stripe excitation.  The solid curve, which is the same in
each of the panels, is for the non-precessing case of electric
field only, with neither strain nor magnetic field.  The dashed
lines in the upper row show the case of applied magnetic field.
The dashed lines in the center row show the case of applied
strain, and the dashed lines in the lower row show the case of
applied strain for strictly 1-D motion. If there is no strain,
Eqs. (4)-(6) for the strictly 1-D motion case are the same as the
2-D case. The values of magnetic field and strain were chosen so
that the spatial precession periods are approximately equal for
the two cases. The values of strain and magnetic field in the
right column of Fig. 3 are twice that in the left column.
Comparing the solid and dashed lines in the upper row of Fig. 3
shows that $n_z$ decreases significantly when a magnetic field is
applied. When the magnetic field is increased, the decay length is
shorter. When strain is applied instead of magnetic field, as in
the center row of Fig. 3, $n_z$ also decays more rapidly. However,
the decay length remains longer for the case of strain than for
the case of an equivalent magnetic field. When strain is applied
for the case of strictly 1-D motion, the decay length is
unchanged.  The accumulated spin-precession angle due to strain
depends only on the electron's location and not on the path taken
to reach that location, resulting in no dephasing of the ensemble
spin-polarization. This behavior does not occur for strictly 1-D
motion with magnetic field.
\begin{figure}
  \includegraphics[angle=0,width=.47\textwidth]{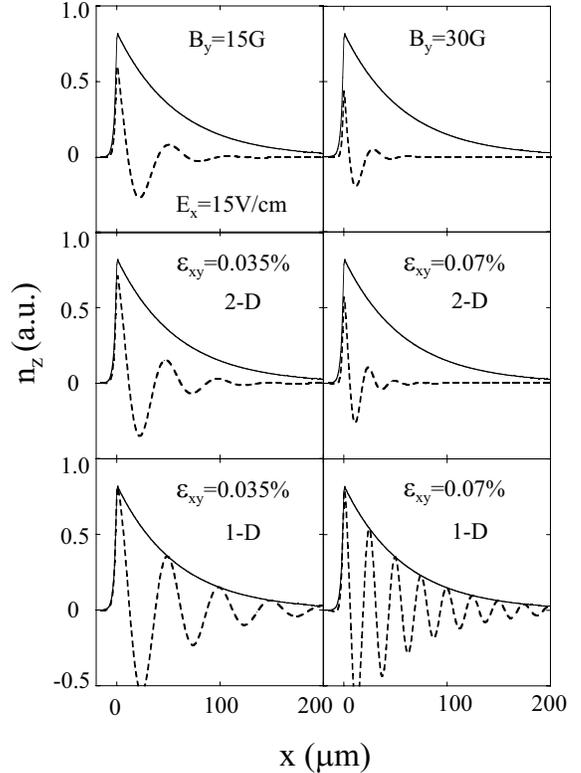}
  \caption{Calculated z-component of spin polarization ($n_z$) generated by a narrow
stripe excitation at $x = 0$ under the influence of a fixed
electric field of $E_x = 15$ V/cm and: Top row, a magnetic field
of $B_y = 15$ G (left) or $B_y = 30$ G (right); center and lower
rows, strain $\epsilon_{xy} = 3.5 \times 10^{-4}$ (left) or
$\epsilon_{xy} = 7 \times 10^{-4}$ (right).  The solid line is for
electric field alone, without magnetic field or strain. The dashed
lines represent cases where, in addition to the electric field, a
magnetic or strain field is applied. The bottom row is for the
case of strain and strictly 1-D motion.  $D=10$ cm$^2$/s,
$\mu=3000$ cm$^2$/Vs, and $\tau_s=125$ ns.
 \label{Fig3}}
\end{figure}

In Fig. 4, we compare the effect of increasing the electric field
while in the presence of either a fixed magnetic field or a fixed
off-diagonal strain. The geometry and material parameters are the
same as for Fig. 3. The two panels in the left column show
calculated results for $B_y = 15$ G and $E_x=10$ V/cm (top panel)
and $E_x=20$ V/cm (bottom panel). The two panels in the right
column show calculated results for a strain of $\epsilon_{xy} =
3.5 \times 10^{-4}$ and the same electric fields as in the left
column. Comparing the top and bottom panels in the left column, we
see that for fixed applied magnetic field the spatial precession
period increases with increasing electric field.  By contrast,
comparing the top and bottom panels in the right column, we see
that for fixed strain the spatial precession period does not
change with increasing electric field.  This behavior was observed
in Ref. \cite{ScottDarryl}. For both strain and magnetic field,
the spatial decay length increases as the electric field
increases, due to the fact that spin transport is governed more by
drift than by diffusion.
\begin{figure}
 \begin{center}
 \epsfig{file=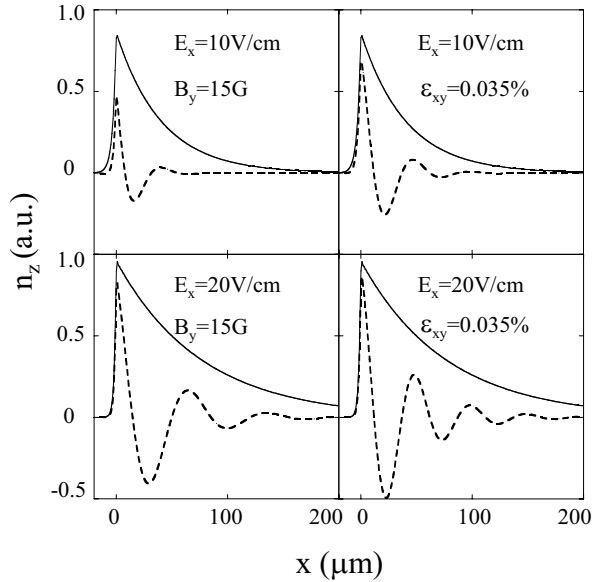,width=0.45\textwidth,clip=,angle=0}
 \end{center}
\caption{Calculated z-component of spin polarization ($n_z$)
generated by a narrow stripe excitation at $x = 0$ under the
influence of a lateral electric field $E_x = 10$ V/cm (top row)
and $E_x = 20$ V/cm (bottom row). $B_y = 15$ G in the left column;
$\epsilon_{xy} = 3.5 \times 10^{-4}$ in the right column.  The
solid line is for electric field alone, without magnetic field or
strain (no precession).  The dashed lines represent cases where,
in addition to the electric field, a magnetic field or strain is
present. $D=10$ cm$^2$/s, $\mu=3000$ cm$^2$/Vs, and $\tau_s=125$
ns.
 \label{Fig4}}
 \end{figure}

\section{Predictions of the Spin-Drift-Diffusion Model}

In this section we present specific examples and predictions of
the spin-drift-diffusion model. We first consider geometries that
correspond to optical spin injection (initial spin density ${\bf
n}_0\parallel \hat{z}$), and then consider geometries that
correspond to electrical spin injection (${\bf n}_0\parallel
\hat{x}$, with spatially extended source contacts).

Figure 5 shows the deleterious effect of diffusion on the spatial
coherence of precessing spin flows. A steady-state spin
polarization ${\bf n}_0 \parallel \hat{z}$ is generated by a
Gaussian source term with 4 $\mu$m full-width at half-maximum
(FWHM), similar to the case of local optical spin injection.
Spin-polarized electrons diffuse, and drift to the right in the
presence of a lateral electric field, $E_x = 10$ V/cm.  We compare
and contrast spin flows in the case of small diffusion constant
(left column; $D=1$ cm$^2$/s) with the case of large diffusion
constant (right column; $D=20$ cm$^2$/s). In Fig. 5(a) and (b),
there is no applied magnetic field and no effective magnetic field
due to strain, therefore the spin density $n_z$ decays
monotonically away from the point of injection. The larger
diffusion constant in Fig. 5(b) is reflected in the greater
diffusive spread of spin polarization along the y-axis.  In Figs.
5(c) and (d) there is an additional magnetic field ($B_y = 20$ G),
and the spins now precess as they flow to the right. With
increasing distance from the point of spin injection, many spin
precession cycles are observed when the diffusion constant is
small (Fig. 5c). In contrast, only a single full precession cycle
is visible when the diffusion constant is large (Fig. 5d).  These
images directly reveal how the effects of an increased diffusion
constant can lead to significant spatial dephasing of precessing
spin flows.  Figures 5(e) and (f) show the spin flow precessing in
the presence of an effective magnetic field due to strain
($\epsilon_{xy} = 7 \times 10^{-4}$).  Again, the degree of
spatial spin coherence is much higher when the diffusion constant
is smaller.  Further, the line-cuts in Figs. 5 (g) and (h) show
that the degree of spatial spin coherence is better maintained for
the case of strain than for the case of magnetic field, as
discussed in section III.

\begin{figure}
 \begin{center}
 \epsfig{file=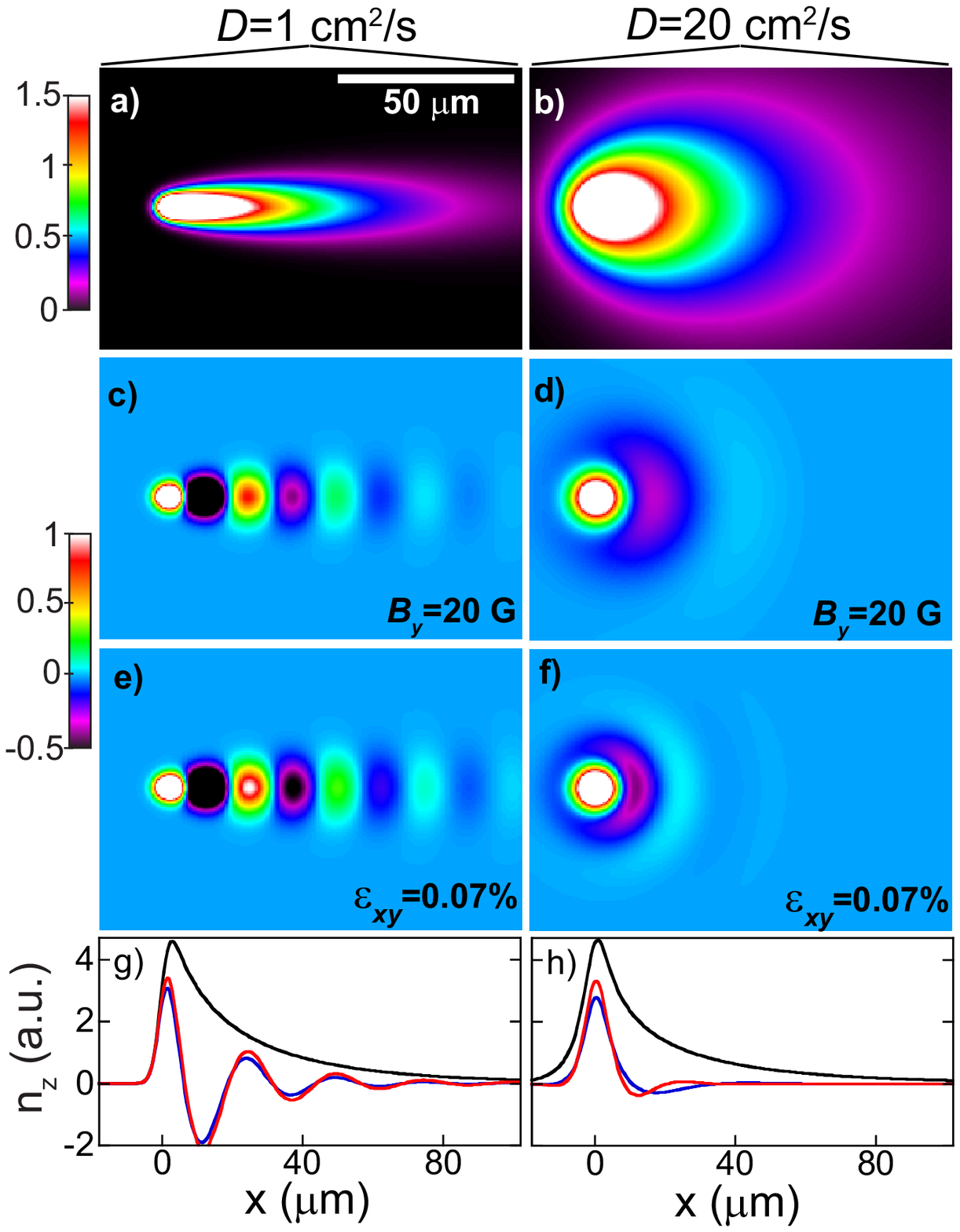,width=0.45\textwidth,clip=}
 \end{center}
\caption{Calculated 80 x 120 $\mu$m images of $n_z$ for
optically-injected spins, showing how diffusion directly
influences the degree of spatial spin coherence for precessing
spin flows. Images in the left column are for $D=1$ cm$^2$/s;
images in the right column are for $D=20$ cm$^2$/s.  All
calculations use $E=10$ V/cm, $\mu=3000$ cm$^2$/Vs, and
$\tau_s=125$ ns. (a) and (b) show spin flow in the absence of
magnetic or strain fields. (c) and (d) show precessing spin flow
in the presence of $B_y = 20$ G. (e) and (f) show precessing spin
flow in the presence of strain ($\epsilon_{xy}=7\times10^{-4}$).
(g) and (h) show line-cuts through the center of the corresponding
images, with blue and red lines showing the case of finite $B_y$
and $\epsilon_{xy}$, respectively. \label{Fig5}}
\end{figure}

Figure 6 shows how the effective magnetic field due to strain
(${\bf B}_\epsilon$) can either augment or oppose a real magnetic
field ($B_y$), leading to markedly different spin flows depending
on the flow direction.  The schematic diagrams at the top of Fig.
6 illustrate how a uniform applied magnetic field $B_y$ is
vectorially summed with the effective magnetic field ${\bf
B}_\epsilon$ due to finite strain, for the case of spin polarized
electrons diffusing radially away from the point of injection (no
drift). Since ${\bf B}_\epsilon$ is always orthogonal to electron
momentum {\bf k}, ${\bf B}_\epsilon$ is chiral for
radially-diffusing spins, as shown. Under the conditions of Fig.
6, the vector sum of these two fields is given by ${\bf B}_{net}$,
which is finite for spins diffusing to the left, and negligible
for spins diffusing to the right. In Fig. 6(a), spin polarization
${\bf n}_0
\parallel \hat{z}$ is generated by a 4 $\mu$m Gaussian FWHM source term, the applied field $B_y = 6$ G,
and the strain is $\epsilon_{xy} = 2 \times 10^{-4}$. Spin
polarized electrons are subject to diffusion only ($E_x = 0$). The
resulting image of $n_z$ is completely asymmetric -- spins
diffusing to the left ``see" a net magnetic field and precess,
while electrons diffusing to the right experience negligible net
magnetic field and do not precess. This behavior was observed in
Ref. \cite{ScottDarryl}. The asymmetry is even more apparent when
the spins drift in the presence of a lateral electric field ${\bf
E}
\parallel \pm \hat{x}$. Fig. 6 (b) shows that spins flowing to the
left ($E_x = -10$ V/cm) exhibit pronounced precession with many
precession cycles, while Fig. 6 (c) shows that spins flowing to
the right under the opposite lateral bias ($E_x = +10$ V/cm) do
not precess ($B_y$ and ${\bf B}_\epsilon$ effectively cancel in
this direction, under these conditions). Fig. 6 (d) shows
corresponding line-cuts through the images.

\begin{figure}
 \begin{center}
 \epsfig{file=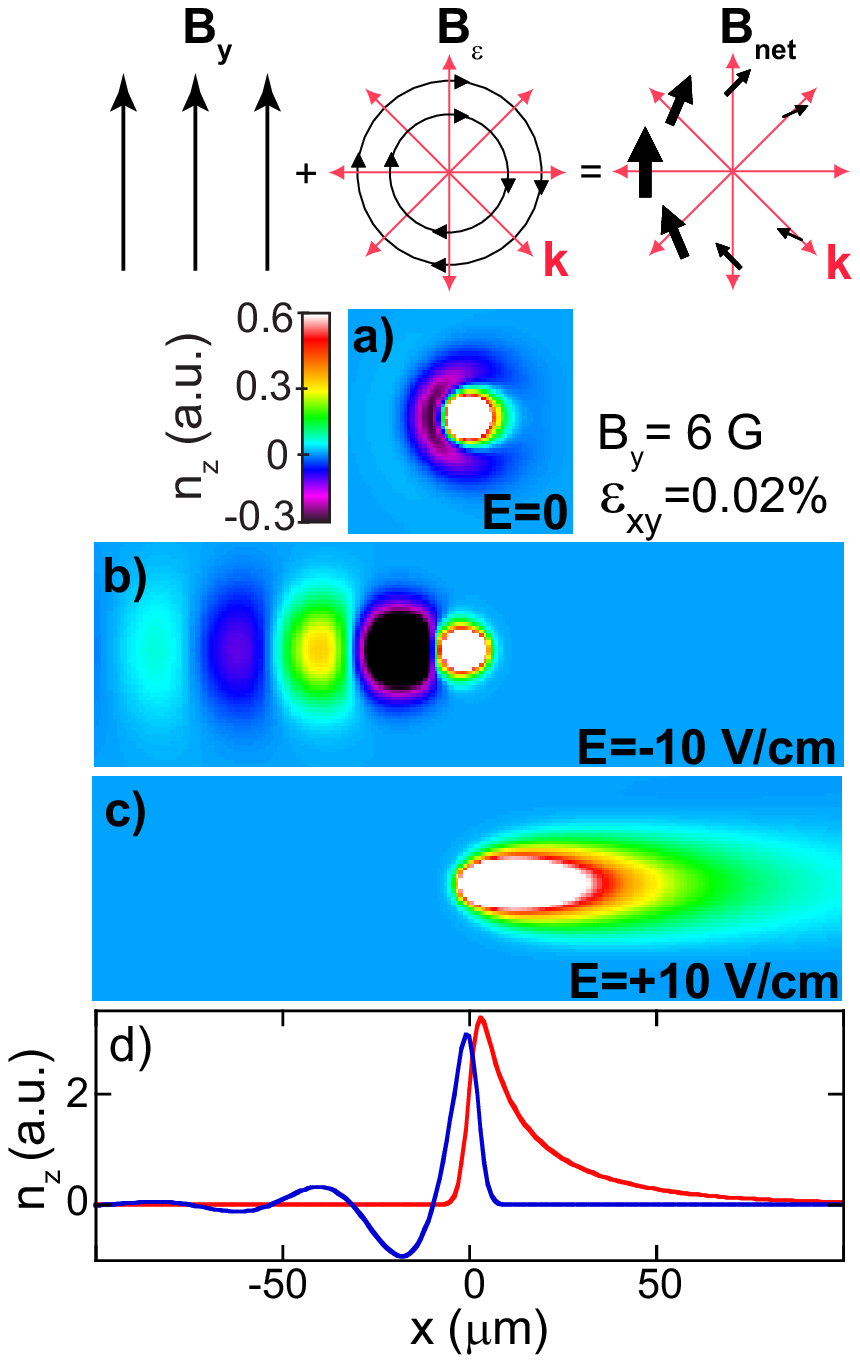,width=0.4\textwidth,clip=}
 \end{center}
\caption{Diagram and calculated images showing how an effective
magnetic field due to off-diagonal strain (${\bf B}_\epsilon$) can
either augment or oppose a real magnetic field $B_y$, depending on
the electron momentum {\bf k}. (a) Calculated 60 x 60 $\mu$m image
of spin diffusion with no electric field, and with $B_y=6$ G and
strain $\epsilon_{xy}=2 \times 10^{-4}$. Spins diffuse radially
away from the point of injection, ``sampling" all momenta {\bf k}
in the x-y plane. Spins diffusing to the left see a net effective
magnetic field and precess; spins diffusing to the right see
negligible net effective magnetic field and do not precess. (b)
With $E_x=-10$ V/cm, electrons flowing to the left show pronounced
spatial precession. (c) With $E_x = +10$ V/cm, electrons flowing
to the right show essentially no precession. (d) Corresponding
line-cuts through the center of the images in (b) and (c). In all
images, $D=3$ cm$^2$/s, $\tau_s = 125$ ns, and $\mu=3000$
cm$^2$/Vs. \label{Fig6}}
\end{figure}

The spin drift-diffusion equations [Eqs. (15) - (18)] can also be
used to compute the in-plane components of electron spin
polarization ($n_x$ and $n_y$).  Knowledge of $n_x$ and $n_y$ is
useful for comparison with experiments that are directly sensitive
to in-plane components of electron spins, such as longitudinal or
transverse (as opposed to polar) magneto-optical Kerr effect
studies, or for direct electrical measurement of in-plane electron
spin using ferromagnetic electrical drain contacts that exhibit a
spin-sensitive conductance \cite{science}.  Figure 7 demonstrates
the utility of Eqs. (15)-(18) by computing all three components of
electron spin polarization ($n_z$, $n_x$, and $n_y$) for the case
of optically injected electrons (${\bf n}_0
\parallel \hat{z}$) diffusing in the presence of both a real uniform magnetic field $B_y$ (left column),
and in the presence of an effective chiral magnetic field due to
strain ${\bf B}_\epsilon$ (middle column). The rightmost column
shows horizontal line-cuts through the center of the corresponding
images of $n_z$ and $n_x$.  The diagrams at the bottom of Fig. 7
illustrate how electron spins, generated at the center of the
image, precess from initially out-of-plane (${\bf n}_0 \parallel
\hat{z}$) to an in-plane direction as they diffuse. In the case of
a uniform applied magnetic field $B_y$, all electrons precess in
the same direction about $B_y$ regardless of their momentum, so
that $n_x$ is finite with circular symmetry, and $n_y$ is
everywhere zero. In contrast, electrons diffusing in the presence
of strain precess about the chiral effective magnetic field ${\bf
B}_\epsilon$ such that their spins are oriented radially away from
the point of injection (at some characteristic radial distance).
This radial spin distribution is reflected in the asymmetric
images of $n_x$ and $n_y$. In the presence of strain, the fact
that the sign of electron spin polarization (for a spin component
orthogonal to the spin generation term) is dependent on the
direction of spin flow has been shown to be a valuable diagnostic
tool (see Ref. \cite{science}).

\begin{figure}
 \begin{center}
 \epsfig{file=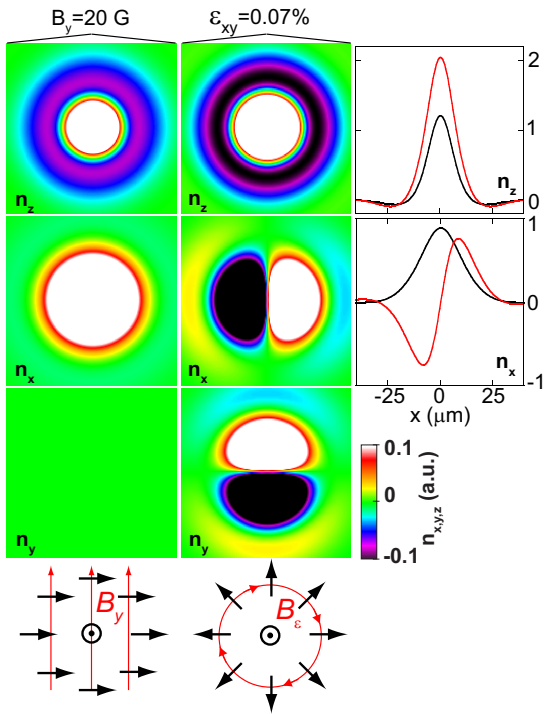,width=0.45\textwidth,clip=}
 \end{center}
\caption{40 x 40 $\mu$m images of the calculated z-, x-, and
y-components of optically-injected spin polarization ($n_z, n_x,
n_y$). Spins are injected with ${\bf n}_0 \parallel \hat{z}$ at
the center of the images, and subject to diffusion only (no
electric field) in the presence of either an applied magnetic
field (left column; $B_y = 20$ G), or strain (middle column;
$\epsilon_{xy} = 7 \times 10^{-4}$). $D=10$ cm$^2$/s and
$\tau_s=125$ ns. The rightmost column shows horizontal line-cuts
through the center of the corresponding images of $n_z$ and $n_x$.
The diagrams at the bottom illustrate how spins precess from
initially out-of-plane (${\bf n}_0 \parallel \hat{z}$) to in-plane
while diffusing in the uniform magnetic field $B_y$, or in the
chiral effective magnetic field ${\bf B}_\epsilon$. \label{Fig7}}
\end{figure}

For electrical spin injection from ferromagnetic contacts, the
generation source is usually spatially extended and the
orientation of injected spins follows the contact magnetization,
which is typically in the x-y plane of the sample surface.
However, polar magneto-optical Kerr microscopy is sensitive to
$n_z$, the z-component of spin polarization.  Therefore, a
magnetic or strain field must be used to rotate the injected spin
polarization out-of-plane in order to allow detection.  In Fig. 8,
we show calculated images of $n_z$ resulting from a (theoretical)
40 $\mu$m wide ferromagnetic stripe contact having in-plane
magnetization ({\bf M}$\parallel \hat{x}$). This spin polarization
source is analogous to the Fe/GaAs spin injection contacts used
experimentally in Ref. \cite{science}. The left column shows $100
\times 200$ $\mu$m images of $n_z$, and the right column shows
horizontal line-cuts through the images.  In Fig. 8(a), electrons
injected with initial spin orientation ${\bf n}_0 \parallel
\hat{x}$ diffuse in the presence of strain ($\epsilon_{xy} = 1
\times 10^{-4}$). Spins diffusing to the left and right precess
into-plane and out-of-plane, respectively, giving the asymmetric
image shown. The maxima and minima of $n_z$ nearly coincide with
the edges of the source contact because the contact width greatly
exceeds the characteristic spin diffusion length $\sqrt{D
\tau_s}$. In Fig. 8(b), we consider the injected spins diffusing
in the presence of a applied magnetic field ($B_y = 1$ G). The
image of $n_z$ is symmetric, because electrons precess in the same
direction (out-of-plane) regardless of which way they diffuse.
Lastly, Fig. 8(c) includes an additional uniform drift electric
field ($E_x = 4$ V/cm), which shifts the spin polarization in the
direction of electron drift (we consider the simplest case here,
in which electric field is uniform across the image; i.e., the
electric field is not generated by the injection contact, nor does
the injection contact distort the uniform electric field.) Note
that finite spin polarization $n_z$ still exists on the
``upstream" side of the source contact due to diffusion of
injected electrons against the net electron current.

\begin{figure}
 \begin{center}
 \epsfig{file=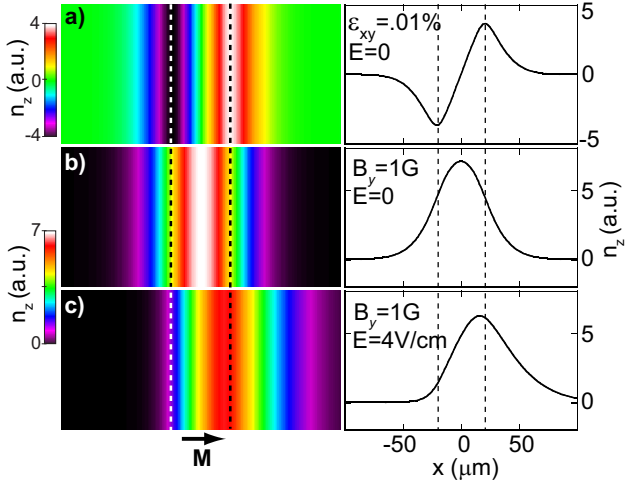,width=0.5\textwidth,clip=}
 \end{center}
\caption{100 x 200 $\mu$m images of the calculated z-component of
electron spin polarization ($n_z$) resulting from {\it in-plane}
spin injection (${\bf n}_0 \parallel {\bf M} \parallel \hat{x}$)
by a 40-$\mu$m wide magnetized stripe contact (edges of the stripe
are shown by dashed lines). Corresponding horizontal line-cuts
through the center of the images are shown to the right. (a) Spin
diffusion alone for the case of a strained sample ($\epsilon_{xy}
= 1 \times 10^{-4}$). Spins diffusing to the left or right precess
in opposite directions. (b) Spin diffusion alone for the case of
an applied magnetic field ($B_y = 1$ G). (c) Spin diffusion and
drift for the case of applied magnetic and lateral electric fields
($B_y = 1$ G, $E_x = 4$ V/cm). For all images, $\mu=3000$
cm$^2$/Vs, $D=10$ cm$^2$/s, $\tau_s=125$ ns. \label{Fig8}}
\end{figure}

Finally, we demonstrate the utility of the spin drift-diffusion
equations in modeling the two-dimensional spin flows resulting
from nontrivially-shaped injection sources (having either in-plane
{\it or} out-of-plane magnetization), such as might be realized in
actual experiments. Figure 9 shows calculated images of $n_z$ in a
semiconductor epilayer resulting from a uniform square
$80\times80$ $\mu$m injection contact. In Fig. 9 (a) and (c), the
contact magnetization (and injected spin polarization) is oriented
out-of-plane (${\bf M} \parallel {\bf n}_0 \parallel \hat{z}$).
For comparison, the contact magnetization (and injected spin
polarization) in Figs. 9 (b) and (d) is in-plane and along the
x-axis (${\bf M} \parallel {\bf n}_0 \parallel \hat{x}$). Spins
drift to the right in an electric field of $E_x = 8$ V/cm in all
of the images.

In actual experiments, it is often desirable to measure the spin
polarization $n_z$ at a specific location on the sample as a
function of in-plane magnetic field $B_y$ \cite{Stephens,science}.
These ``Hanle curves" \cite{Meier} provide valuable information on
the dynamics of the flowing spins, and in certain cases can be
used to infer spin drift velocities and also, in conjunction with
controlled strain, the direction of spin flow \cite {science}.
Figures 9(e) and (f) show calculated Hanle curves acquired at 4,
20, 40, and 80 $\mu$m from the right edge of the square injection
contact. Solid (dashed) lines show the Hanle curves in the absence
(presence) of strain. For the case of out-of-plane injection [Fig.
9 (e)], the Hanle curves are symmetric with respect to $B_y$. In
contrast, for the case of in-plane injection [Fig. 9(f)], the
Hanle curves are antisymmetric with respect to $B_y$, as expected
for orthogonal spin generation and detection axes. Unlike
conventional Hanle measurements which are typically sensitive to
the spin polarization of {\it all} injected spins (e.g., the case
of optical spin injection and detection via spatially-integrated
luminescence polarization), in these studies the injection source
and the point of detection may be spatially separated, so that
only a subset of the injected electrons are detected by the probe
laser. Further, these experiments allow access to a regime wherein
the spin drift length exceeds the characteristic spin diffusion
length. In this regime the Hanle curves can exhibit highly
atypical shapes, even showing multiple oscillations as $B_y$ is
swept \cite{science}. For the images of Fig. 9, where the injected
spins subsequently flow to the right, the effect of strain is to
shift the Hanle curves to negative values of $B_y$ (dashed lines),
implying that the effective magnetic field due to strain (${\bf
B}_\epsilon$) is oriented along $+\hat{y}$. Conversely, a shifted
Hanle curve can be used to infer the presence of strain in the
sample, and also can be used (in conjunction with controlled
strain) to infer the actual direction in which the spin polarized
electrons are flowing \cite{science}.

\begin{figure}
 \begin{center}
 \epsfig{file=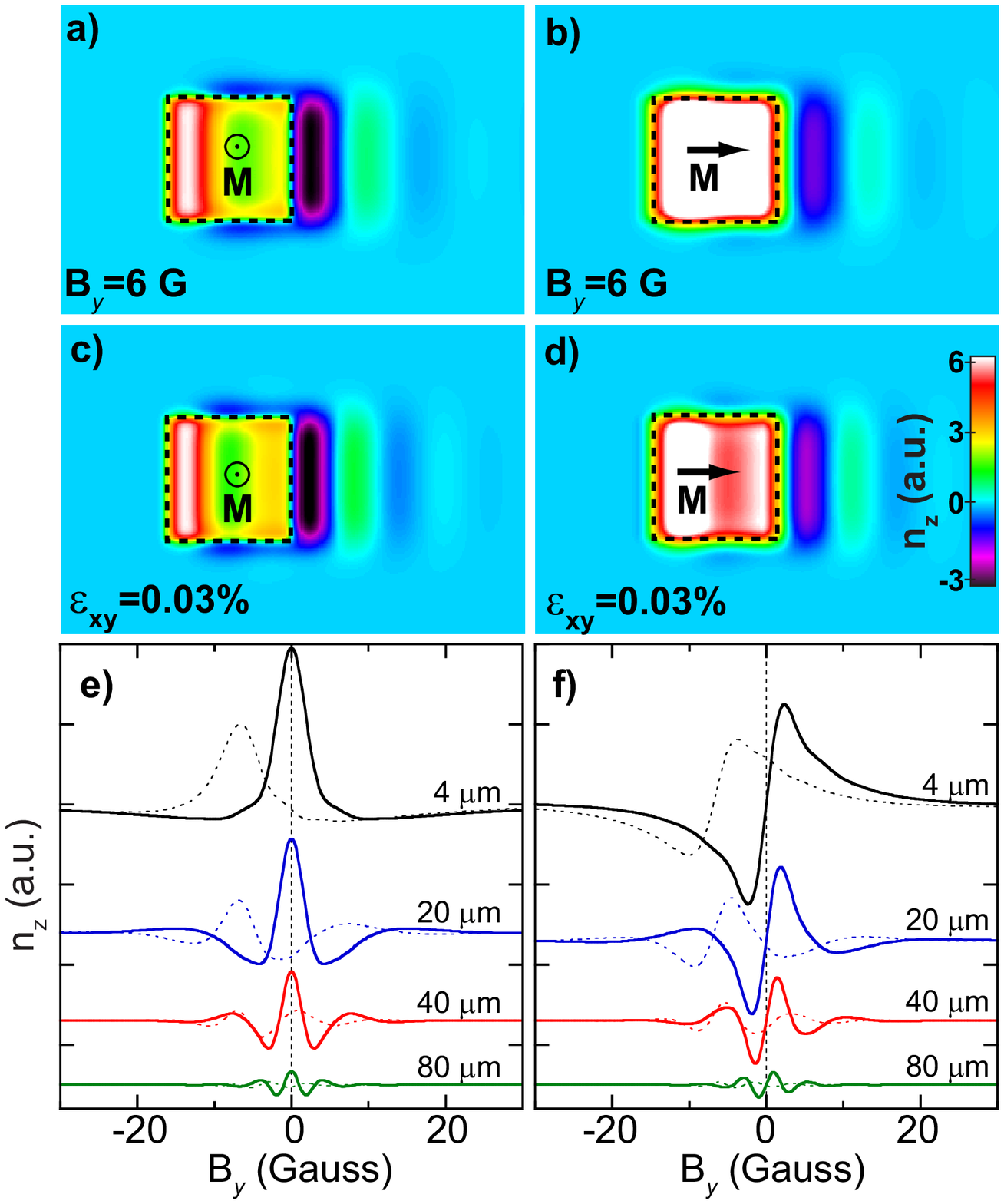,width=0.5\textwidth,clip=}
 \end{center}
\caption{Calculated images of $n_z$ resulting from either
out-of-plane (left column) or in-plane (right column) spin
injection and subject to either magnetic or strain fields. The 300
x 200 $\mu$m images show the z-component of calculated electron
spin polarization ($n_z$) resulting from a square 80 x 80 $\mu$m
injection contact (edges of the source are indicated by dashed
lines). In (a) and (c) the contact magnetization is out-of-plane
(${\bf n}_0 \parallel \hat{z}$), and in (b) and (d) the contact
magnetization is in-plane (${\bf n}_0 \parallel \hat{x}$).
Measuring $n_z$ at a specific position as a function of in-plane
magnetic field $B_y$ yields the ``Hanle curves" [40] 
shown
in (e) and (f). These curves are acquired at distances of 4, 20,
40 and 80 $\mu$m from the right edge of the square source contact.
Solid lines correspond to sweeping $B_y$ alone; dashed lines
include the presence of strain ($\epsilon_{xy}=0.03\%$), which
shifts the Hanle curves to the left.  In the calculations, $\mu =
3000$ cm$^2$/Vs, $E_x=8$ V/cm, $D=3$ cm$^2$/s, and $\tau_s = 125$
ns. \label{Fig9}}
\end{figure}

 \section{Summary and Conclusion}

We investigated the effects of electric fields, magnetic fields,
and off-diagonal strain on the transport of spin-polarized
electrons in zincblende semiconductors. Starting with a quantum
kinetic approach, we first derived a semiclassical kinetic theory
of electron spin dynamics and transport, and from this kinetic
theory constructed spin-drift-diffusion equations when the total
electron density is position independent and for spatially uniform
electric, magnetic, and strain fields. This case of spatially
uniform fields and uniform electron density corresponds to the
experimental situation realized in Refs. \cite{ScottDarryl} and
\cite{science}. We compared the results of our
spin-drift-diffusion model with Kerr microscopy images and found
very good agreement.  Our semiclassical kinetic theory is
formulated on a microscopic foundation and can be extended to the
cases of spatially varying (electric, magnetic and strain) fields
and electron density.  We contrasted the spin precession resulting
from magnetic and strain fields.  We found that because
spin-strain coupling depends linearly on electron wave vector
whereas spin-magnetic field coupling is independent of electron
wave vector {\bf k}, spatial coherence of electron spin precession
is much better maintained with strain than with magnetic fields.
This effect is most dramatic for 1-D systems. In contrast to the
case of a magnetic field, we find that the spatial period of
electron spin precession is independent of the applied electrical
bias in samples with off-diagonal strain. The freedom to operate
at variable electric bias may benefit functional devices that are
based on rotation of spin from one point in space to the other, as
in spin transistor designs. Finally, we explore the utility of the
spin drift-diffusion model by considering spin flows in a variety
of potential cases, such as in-plane and out-of-plane spin
injection from spatially extended sources in the presence of
electric, magnetic, and/or strain fields.

\section{Acknowledgment}
The authors thank Lev Boulaevskii, Thomas Luu, Slobodan Mati\'{c},
Paul Crowell, Chris Palmstr{\o}m and Peter Littlewood for useful
discussions and input. This work was supported by the Los Alamos
National Laboratory LDRD program.

\section{Appendix: Semiclassical Kinetic Theory of Electron Spin Dynamics with
Electric, Magnetic and Strain Fields}

We start from the Keldysh Green's function description
\cite{Keldysh} for a system driven out of equilibrium
\cite{Kuznetsov}. The Dyson equation
\begin{equation}
(\hat {G} _0 ^{-1} - \hat {\Sigma}) \otimes \hat G = \delta (1,2)
\end{equation}
is written for the 4x4 matrix Green's function in the tensor
product of spin and Keldysh space
\begin{equation}
\hat G=\left( \begin{array}{cc}
G^R & G^K \\
0 & G^A
\end{array}
\right) ,
\end{equation}
 where $G^R$, $G^A$ and $G^K$ are the retarded, advanced and Keldysh
 Green's function which are 2x2 matrices in spin space
 (for more details on this Larkin-Ovchinnikov notation see Ref.
 \cite{RammerSmith}).  The operator $\hat G _0 ^{-1} $ is defined as
 \begin{eqnarray}
\hat G _0 ^{-1} (1,2) &=& \left( \begin{array}{cc}
G_0^{-1} & 0 \\
0 & G_0^{-1}
\end{array}
\right) \ , \nonumber\\
G_0^{-1} &=& [i \hbar \partial _{t_1} - \hat {\epsilon} (1)]
\delta (1,2),
\end{eqnarray}
where the indices $1,2$ represent the first and the second set of
coordinate, time and spin variables and $\delta (1,2)$ represents
a product of delta-functions in these variables; the operation
$\otimes$ denotes matrix product in relevant spaces.  Here $\hat
{\Sigma}$ is the self-energy operator while $\hat {\epsilon}$
expresses the effective mass Hamiltonian for the two spin states
near the conduction band minimum. It is diagonal in Keldysh space
and given by \cite{Meier}
\begin{eqnarray}
\hat {\epsilon} &=&\left( \begin{array}{cc}
\epsilon & 0 \\
0 & \epsilon
\end{array}
\right) \ , \nonumber\\
\epsilon (1) &=& \frac{1}{2 m} [ {\bf p}_{\bf {x_1}}- \frac{e}{c}
{\bf A} (1) ]^2 +e \phi (1) + \frac{\hbar}{2}  \mbox {\boldmath
$\Omega$ }^{eff} \cdot \mbox {\boldmath $\sigma$},
\nonumber\\
 \mbox {$\hbar$ \boldmath $\Omega$} ^{eff}&=&  g \mu _B {\bf B}
+  C_3 \mbox {\boldmath $ \varphi$}  + \frac{\alpha _c \hbar ^3}{
\sqrt{2 m^3 E_g}} \mbox {\boldmath $\kappa$} +\alpha
(\mbox{\boldmath{$k$}}\times \mbox{\boldmath{$\hat{n}$}}) \ .
\nonumber
\end{eqnarray}

The last term in $\mbox {\boldmath $\Omega$} ^{eff}$ with the
coefficient $\alpha$ is the Rashba term (for an asymmetric
heterostructure with symmetry-breaking axis $\bf \hat{n}$), while
\begin{equation}
\kappa _x=  k_x (k_y ^2 - k_z^2)
\end{equation}
is the Dresselhaus term arising from bulk inversion asymmetry,
$\alpha _c = (4/3) (m/m_{cv}) \eta/\sqrt{1-\eta /3}$, $m_{cv}$
determines an interband term quadratic in $\bf k$ and is related
to the interaction of the conduction and valence bands with other
bands.  The components $\kappa_y$ and $\kappa_z$ are obtained by
cyclic permutations of Eq. (26). From symmetry considerations
alone \cite{Meier,Bernevig} $\mbox {\boldmath $\Omega$} ^{eff}$
could also contain terms that couple spin to {\it diagonal} strain
components, e.g., terms of the form $\xi _z=k_z (\epsilon
_{xx}-\epsilon _{yy})$, etc. However, these terms only appear if
spin-orbit mixing of states widely separated in energy is included
[see Ref.\cite{Meier} for more details] and therefore their
numerical coefficients are small.

The Dyson equation gives an exact description of the system.
However, it cannot be reduced to an expression involving
equal-time Green's functions which determine physical quantities,
such as densities and particle currents, via the single-particle
density matrix. Deriving the quantum-kinetic equation involves
considering the difference of the Dyson equation and its
conjugate,
\begin{equation}\label{DysonMinusConj}
[\hat G_0 ^{-1} - \hat {\Sigma} \stackrel{\otimes}{,} \hat G]_- =
0
\end{equation}
and approximations including a gradient expansion and a
quasiclassical approximation. (Here $[A \stackrel{\otimes}{,}
B]_-$ denotes a commutator of quantities $A$ and $B$ with respect
to the operation $\otimes$.) We use a result due to Langreth
\cite{Langreth} within the gradient expansion to obtain the
kinetic equation for a gauge invariant distribution function.

To introduce the gradient expansion, we transform to a mixed
Wigner representation:
\begin{eqnarray*}
& {\bf R}=\frac{{\bf r_1}+{\bf r_2}}{2}, \ T=\frac{t_1+t_2}{2}, \
{\bf r} = {\bf r_1}-{\bf r_2}, \ t=t_1-t_2 & \ .
\end{eqnarray*}
After performing a Fourier transform with respect to relative
coordinates $\bf r$ and $t$
\begin{equation}
\hat G (X,p)=\int dt \int d{\bf r} e^{-i p \cdot x } \hat
G(X+x/2,X-x/2)\ ,
\end{equation}
where $X=(T,{\bf R})$, $x=(t, {\bf r})$, $p=(\omega,{\bf
p}/\hbar)$ and the dot product is given by $p \cdot x = -\omega t
+ {\bf p} \cdot {\bf r}/\hbar $, we represent the Fourier
transform as a Taylor expansion expression involving products of
those quantities in the Fourier transformed mixed representation:
\begin{equation}\label{GradExp}
(M \otimes N)(X,p)= e^{i   (\partial _X^M \cdot \partial _p^N -
\partial _p^M \cdot \partial _X^N)/2} M(X,p) N(X,p) \ .
\end{equation}
We defined the exclusive partial derivatives $\partial
_X^M=(-\partial _T, \nabla _{\bf R})$, $\partial _p ^M =
(-\partial _{\omega},\hbar \partial _{\bf p})$ acting only on the
quantity M; the scalar product of derivatives amounts to
 $\partial _X \cdot \partial _p = - \partial _T \partial _{\omega} + \hbar \partial _{\bf R}
 \partial _{\bf p}$.
In the mixed representation the operator $ G_0^{-1}$ is given by
\begin{eqnarray}
G_0^{-1} ({\bf R};\omega,{\bf p}) &=& \hbar \omega - \frac{1}{2 m}
({\bf p}- \frac{e}{c} {\bf A})^2
-e \phi ({\bf R}) \nonumber\\
& &-\frac{\hbar}{2}  \mbox{ \boldmath $\Omega$} ^{eff}  ({\bf
R};{\bf p}) \cdot \mbox{ \boldmath $\sigma$} \ .
\end{eqnarray}
Using the gradient expansion Eq.~(\ref{GradExp}),
Eq.~(\ref{DysonMinusConj}) is written as
\begin{eqnarray}
& 0=[\hat G_0^{-1}-\hat {\Sigma}\stackrel{\otimes}{,} \hat
G]_-(T,{\bf R};\omega,{\bf p}) =   i  [\hat G_0^{-1} - \hat
{\Sigma} , \hat G ]_p  - &
    \nonumber\\
& [\frac{\hbar}{2} \mbox {\boldmath $\Omega $} ^{eff}  ({\bf
R};{\bf p}) \cdot \sigma  \odot \hat 1 + \hat {\Sigma} (T,{\bf
R};\omega,{\bf p}), \hat G  (T,{\bf R};\omega,{\bf p})]_- , &
\end{eqnarray}
where
\begin{eqnarray}
[M,N]_p &=& \frac{1}{2} (\partial _X^M M \partial _p ^N N
- \partial _p^M M \partial _X^N N)  \nonumber\\
&& - \frac{1}{2} (\partial _X^N N \partial _p ^M M - \partial _p^N
N \partial _X^M M)
\end{eqnarray}
is the generalized Poisson bracket and $\odot$ denotes a tensor
product.

Within the gradient approximation, we introduce gauge-invariant
quantities as
\begin{equation}
{\tilde A}(\Omega,{\bf P},{\tilde {\bf R}}, {\tilde T}) = A(
\omega, {\bf p}, {\bf R}, T)
\end{equation}
by a change of variables
\begin{equation}
{\bf P} = {\bf p} - \frac{e {\bf A}}{c} ({\bf R},T), \ \Omega = \omega -
(e/\hbar) \phi ({\bf R},T) \ .
\end{equation}
The relations between the derivatives in old and new coordinates
are
\begin{eqnarray*}
\partial _{\bf R} &=& -\frac{e}{\hbar} \nabla \phi \partial _{\Omega}
-\frac{e}{c} \sum _{i,j} \hat e_j\frac{\partial A_i}{\partial R_j}  \partial
_{ P_i}
+ \partial _{\tilde {\bf R}} , \\
\partial _T &=&-e \frac{\partial \phi}{\partial T} \partial _{\Omega}
-\frac{e}{c} \frac{\partial {\bf A}}{\partial T} \cdot \partial _{\bf P}
+\partial _{\tilde T} ,
\end{eqnarray*}
and $\partial _{\bf p}=\partial _{\bf P},\ \partial _{\omega}=
\partial _{\Omega}$, where ${\bf E}$ is the electric field. From
these relations one can prove the identity connecting the
generalized Poisson brackets in old and new coordinates
\cite{Langreth}
\begin{eqnarray}
&[M,N]_p = [{\tilde {M}},{\tilde {N}}]_P + & \nonumber\\
 &  \frac{e}{2} \{ [
 {\bf B} \cdot (\nabla _{\bf P}^{\tilde M} {\tilde M}
\times \nabla _{\bf P} ^{\tilde N} {\tilde N} )
 +      {\bf E} \cdot
(\partial _{\Omega}^{\tilde M} {\tilde M} \nabla _{\bf P}^{\tilde
N} {\tilde N} -  \nabla _{\bf P}^{\tilde M}{\tilde M} \partial
_{\Omega}^{\tilde N}
{\tilde N}) ] & \nonumber\\
& - [N \leftrightarrow M] \} \ .& \nonumber
\end{eqnarray}
The generalized Poisson bracket in the new variables is given by
\begin{eqnarray}
&\frac{1}{\hbar} [ \hat G_0^{-1},\hat G]_P = \partial _T \hat G +
{\bf v} \cdot \partial _{\bf R} \hat G
& \\
& + \frac{\hbar }{2} \sum _{i,j} \sigma _i ([\partial _{P_j} \Omega _i
^{eff} ,
\partial  _{ R_j} \hat G]_+
- [ \partial _{ R_j} \Omega _i ^{eff} ,
\partial  _{ P_j} \hat  G]_+ ) , & \nonumber
\end{eqnarray}
where ${\bf v} = {\bf P}/m$ and $[A,B]_+$ denotes an
anticommutator. The Dyson equation in gauge invariant form becomes
\begin{eqnarray}
&
  \{ \partial _T  + {\bf v} \cdot \partial _{\bf R}
 + e  {\bf E} \cdot ( \partial _{\bf P}  + \frac{{\bf v}^{eff}}{\hbar}
\partial _{\Omega} )
\} \hat G
&   \nonumber\\
& + \frac{1 }{2} \sum _i  ([\sigma _i\partial _{\bf P} \Omega _i
^{eff} ,
\partial  _{\bf R} \hat G]_+
-[ \sigma _i \partial _{\bf R} \Omega _i ^{eff} ,
\partial  _{\bf P} \hat G]_+)
&  \nonumber\\
& + \frac{e}{c}({\bf v}^{eff} \times {\bf B}) \cdot \partial _{\bf P} \hat G
+\frac{i}{2}[\mbox {\boldmath $\Omega $}^{eff} \cdot \mbox
{\boldmath $\sigma$}, \hat G]_- = -i [\hat {\Sigma}, \hat G]_- , &
\label{KinEqGOut}
\end{eqnarray}
where ${\bf v}^{eff}={\bf v}-(\hbar /2)\sigma _i \partial _{\bf P}
\Omega _i ^{eff}$. Equation ~(\ref{KinEqGOut}) is a central result
of our microscopic derivation. It describes spin dynamics in the
presence of electric, magnetic and strain fields provided the
lengthscale of spatial variation is large compared to the
characteristic lengthscale for wavefunction variation. For the
case of uniform strain, the strain terms in Eq.~(\ref{KinEqGOut})
agree with those found in Ref.\cite{Mishchenko}. In the absence of
strain fields, the half-sum of the $R$, $A$ and $K$ components of
Eq. ~(\ref{KinEqGOut}) in the equal-time limit yields the kinetic
equation obtained in Ref. \cite{Bronold} for the distribution
function $N = -(i/2)(G^K+G^R-G^A)$ \cite{Kuznetsov}.

The diagonal components $G^R$ and $G^A$ of the Green's function
characterize the states, while the Keldysh component $G^K$
contains information on the occupation of these states
\cite{RammerSmith}. We consider the Keldysh component of Eq.
~(\ref{KinEqGOut}) in order to derive the kinetic equation. A
distribution function $ f(T,{\bf R};\Omega,{\bf P})$ is introduced
by the ansatz
 \begin{equation}
 G^K = G^R \otimes  f - f \otimes G^A \ .
 \end{equation}
Using $[G_0^{-1}-\Sigma ^R \stackrel{\otimes}{,} G^R]_-=0$ and
$[G_0^{-1}-\Sigma ^A \stackrel{\otimes}{,} G^A]_-=0$, the Keldysh
component of Eq. ~(\ref{KinEqGOut}) gives
\begin{equation}\label{KinEqB}
G^R \otimes B - B \otimes G^A = 0 \ ,
\end{equation}
where the $2\times2$ matrix $B$ is given by
\begin{eqnarray}
B[ f] &=& [G_0^{-1} \stackrel{\otimes}{,}  f]_-
- \Sigma ^R \otimes  f + f \otimes \Sigma ^A +\Sigma ^K  \nonumber\\
& \equiv & [G_0^{-1} \stackrel{\otimes}{,}  f]_-
 - C(\hat {\Sigma}, f) ,
\end{eqnarray}
and in the mixed representation we have
\begin{eqnarray}
& -\frac{i}{\hbar} [G_0^{-1} \stackrel{\otimes}{,}  f]_- = i
\frac{1}{2} [\mbox {\boldmath $\Omega $} ^{eff} \cdot \mbox
{\boldmath $\sigma $},f]_- &
\nonumber\\
& +\frac{\hbar}{2} \sum _i \left([\sigma _i \partial _{\bf P} \Omega
_i ^{eff} ,
\partial  _{\bf R} f]_+
-[ \sigma _i \partial _{\bf R} \Omega _i ^{eff} ,
\partial  _{\bf P} f]_+\right)   &  \nonumber\\
&+ \{ \partial _T  + {\bf v} \cdot \partial _{\bf R}
 + e  {\bf E} \cdot ( \partial _{\bf P}  + \frac{{\bf v}^{eff}}{\hbar}
\partial _{\Omega} )\} f ,
&
\end{eqnarray}
where the Lorentz force term was neglected. In the geometry that
we consider (Fig. 1), electrons are confined to move in a
two-dimensional layer perpendicular to [001], and electric and
magnetic fields are applied in the plane of electron motion. Hence
the Lorentz force is along the z-axis and does not affect the
motion in this case.

We seek the solution of Eq.~(\ref{KinEqB}) for the distribution
function as the solution of the equation $B[f]=0$ i.e.
\begin{equation}\label{KinEqBh}
[G_0^{-1} \stackrel{\otimes}{,}  f]_-
 = C(\hat {\Sigma}, f) \ .
\end{equation}
Taking the equal-time limit (which amounts to integrating over
$\Omega$ so that terms with derivative $\partial _{\Omega}$
vanish),
  Eq.~(\ref{KinEqBh}) is then written as
\begin{eqnarray}\label{uniformkinetic}
& \{ \partial _T  + {\bf v} \cdot \partial _{\bf R}
 + e  {\bf E} \cdot  \partial _{\bf P}
 \} h
& \nonumber\\
& + \frac{\hbar}{2} \sum _i ([\sigma _i \partial _{\bf P} \Omega _i
^{eff} ,
\partial  _{\bf R} h]_+
-[ \sigma _i \partial _{\bf R} \Omega _i ^{eff} ,
\partial  _{\bf P} h]_+)
&
 \nonumber\\
& +\frac{i}{2}[\mbox {\boldmath $\Omega $} ^{eff} \cdot \mbox
{\boldmath $\sigma $},h]_- = -\frac{i}{\hbar} \int d\Omega C(\hat
{\Sigma},f) ,
\end{eqnarray}
where $h(T,{\bf R},{\bf P})=\int d\Omega f(T,{\bf R};\Omega,{\bf
P})$.

For the experimental situation that we consider the applied
electric, magnetic, and strain fields are spatially uniform and
the net electron density is constrained by electrostatics to be
spatially uniform. It is convenient to write $h={\rho_0 +
\sum_{i=1}^{3}\rho_i\sigma_i}$. For the steady state, spatially
uniform case considered experimentally, Eq.~(\ref{uniformkinetic})
for the spin components of $\mbox {\boldmath $\rho $}$ becomes a
Boltzmann type equation of the form
\begin{eqnarray}\label{BoltzmannEq}
\left\{ \frac{{\bf P}}{m} \cdot \partial _{\bf R}
 + e  {\bf E} \cdot  \partial _{\bf P}
 \right\} {\rho}_i - [\mbox {\boldmath $\Omega $} ^{eff} \times \mbox
{\boldmath $\rho $}]_i = {I^i}_C ,
\end{eqnarray}
where ${I^i}_C$ is a collision integral representing the right
hand side of Eq.~(\ref{uniformkinetic}).  To construct a
spin-drift-diffusion equation from Eq.~(\ref{BoltzmannEq}), we
make the ansatz
\begin{equation}\label{ansatz}
\rho_i ({\bf R}, {\bf P}) = n_i ({\bf R}) f^0 ({\bf P}) + \delta_i
({\bf R}, {\bf P} ) ,
\end{equation}
where $n_i (\bf R)$ is a spin density that is to be determined,
$f^0 (\bf P)$ is a known equilibrium momentum distribution
function and $\delta_i (\bf R, \bf P )$ is a small correction. The
collision integral is described using a relaxation time
approximation
\begin{equation}
{I^i}_C = - \frac {\delta_i (\bf R, \bf P )}{\tau} ,
\end{equation}
where $\tau$ is a momentum scattering time that may depend on $\bf
P$ but not on $\bf R$. Substituting Eq.~(\ref{ansatz}) into
Eq.~(\ref{BoltzmannEq}) and neglecting contributions for the
correction term $\delta_i$ on the left hand-side gives a first
approximation for $\delta_i$
\begin{eqnarray}\label{delta}
\delta_x = - \tau \biggl{[}\left(\frac{\bf P}{m} \cdot
\frac{\partial n_x}{\partial \bf R}\right) f^0 + n_x \left(e {\bf
E} \cdot \frac{\partial f^0}{\partial \bf P}\right) - \nonumber \\
(\Omega^{eff}_y n_z - \Omega^{eff}_z n_y ) f^0 \biggr{]}
\end{eqnarray}
with cyclic permutation of indices for the other components of
$\delta_i$.  This procedure can be iterated to give a systematic
expansion for $\delta_i$, that is $\delta_i = \delta^0_i +
\delta^1_i + ... $ with the momentum scattering time $\tau$ being
the small parameter in the expansion. We consider the case of
rapid momentum scattering and keep only the lowest order term.

We integrate Eq.~(\ref{BoltzmannEq}) over all momenta to give
\begin{eqnarray}\label{integralP}
\partial _{\bf R} \cdot \int d^3 {\bf P} \left(\frac{{\bf P}}{m}{\rho}_i\right)
 + e  {\bf E} \cdot  \int d^3 {\bf P} (\partial _{\bf P} {\rho}_i)
  \nonumber \\ - \int d^3 {\bf P} [\mbox {\boldmath $\Omega $} ^{eff} \times \mbox
{\boldmath $\rho $}]_i =  \int d^3 {\bf P} ({I^i}_C) = (G_i - R_i)
.
\end{eqnarray}
The second term on the left-hand side of Eq.~(\ref{integralP})
vanishes identically and the possibility of spin generation
($G_i$) and relaxation ($R_i$) has been included empirically in
the momentum integration of the scattering integral. [A more
rigorous discussion of how momentum scattering (Elliot-Yafet
mechanism) and the Dresselhaus term in the Hamiltonian
(Dyakonov-Perel mechanism) lead to spin relaxation within this
formal structure is given in Ref.\cite{Bronold}.]  Substituting
Eq.~(\ref{ansatz}) into Eq.~(\ref{integralP}) and using
Eq.~(\ref{delta}) for $\delta_i$, gives
\begin{eqnarray}\label{bigmess}
& -\partial_{\bf R} \cdot \int d^3 {\bf P}\left(\frac{{\bf P}}{m}\right) {\tau}
\left[\left(\frac{{\bf P}}{m} \cdot  \partial _{\bf R} n_x\right)
f^0
 + n_x (e  {\bf E} \cdot  \partial _{\bf P} f^0)\right]
& \nonumber \\ &-\left[ n_z \int d^3 {\bf P} (\Omega^{eff}_y f^0) - n_y \int d^3 {\bf P} (\Omega^{eff}_z f^0)\right]
& \nonumber \\ & + \int d^3 {\bf P} \tau \Omega^{eff}_y\left[\left(2 \frac{{\bf P}}{m} \cdot \partial _{\bf R} n_z\right)f^0
  + n_z (e  {\bf E} \cdot  \partial _{\bf P} f^0)\right]
& \nonumber \\ & - \int d^3 {\bf P} \tau \Omega^{eff}_z\left [\left(2 \frac{{\bf P}}{m} \cdot \partial _{\bf R} n_y\right)f^0
  + n_y (e  {\bf E} \cdot  \partial _{\bf P} f^0)\right]
& \nonumber \\ & + \int d^3 {\bf P} \tau [ (\Omega^{eff}_z)^2  +  (\Omega^{eff}_y)^2] f^0 n_x
& \nonumber \\ &- \int d^3 {\bf P} \tau  \Omega^{eff}_x ( \Omega^{eff}_y  n_y+  \Omega^{eff}_z n_z) f^0  & \nonumber \\ 
& = G_x - \frac{n_x}{\tau_s} &   
\end{eqnarray}
with cyclic permutations of indices for the other components of
$n_i$. Here the spin relaxation was taken to have the form $R_i =
{n_i}/{\tau_s}$ where $\tau_s$ is a spin relaxation time. Using
the form given in Eq. (1) for $\mbox {\boldmath $\Omega $} ^{eff}$
gives Eqs. (4)-(6) with the mobility and diffusivity given by
\begin{eqnarray}\label{mobility}
\mu = e \int d^3 {\bf P} \tau \frac{P^2}{3m^2}
(\partial_{\epsilon} f^0)
\end{eqnarray}
and
\begin{eqnarray}\label{diffusivity}
D = \int d^3 {\bf P} \tau \frac{P^2}{3m^2}  f^0  ,
\end{eqnarray}
where $\partial_{\epsilon}$ denotes a derivative with respect to
energy.  The first term on the left-hand side (LHS) of
Eq.~(\ref{bigmess}) gives the usual diffusion and drift terms. The
second term on the LHS, which does not contain the momentum
scattering time $\tau$, is even in momentum for the magnetic field
contributions to $\mbox {\boldmath $\Omega $} ^{eff}$ and odd in
momentum for the strain contributions.  Thus, the magnetic field
contribution to this term survives the momentum integration but
the strain contribution does not. By contrast the third and fourth
terms on the LHS contribute for the strain terms but not for the
magnetic field terms. The fifth term on the LHS is even in
momentum for both magnetic field and strain terms, but odd for the
cross term. For the strain terms there are quadratic-in-momentum
contributions from $\mbox {\boldmath $\Omega $} ^{eff}$ that
combine with the momentum scattering time $\tau$ to give a
diffusivity whereas the magnetic field terms are proportional to
the averaged momentum scattering time.  Because the momentum
scattering time is short, these magnetic field dependent terms are 
very small and can be neglected. For the experimental geometry 
$\Omega_z^{eff}$ is zero and $\Omega_x^{eff}/P_y=\Omega_y^{eff}/P_x$.  
Therefore, the contribution of this term in the equation for $n_z$ 
is twice that in the corresponding equations for $n_x$ and $n_y$. 
This accounts for the factor of two multiplying $C_s^2D$ in Eq. (6) 
compared to Eqs, (4) and (5).  For strictly 1-D motion in the 
$x$-direction $\Omega_y^{eff}$ vanishes and the factor of two 
multiplying $C_s^2D$ in Eq. (6) is replaced by unity. The sixth 
term on the LHS vanishes for the geometry that we consider.

\end{multicols}

\end{document}